\documentclass[aoas,MSNbibl,nameyear,dvips]{arximspdf}
\usepackage{graphicx}

\doi{10.1214/11-AOAS501}
\volume{6}
\issue{1}
\pubyear{2012}
\firstpage{409}
\lastpage{427}

\makeatletter

\def\bsuffix #1{#1}

\newcommand{\field}[1]{\mathbb{#1}}
\newcommand{\R}{\field{R}}
\newcommand{\p}{\field{P}}
\newcommand{\N}{\field{N}}
\newcommand{\Z}{\field{Z}}
\newcommand{\E}{\field{E}}

\newcommand{\argmin}{\mathop{\arg\min}}

\newtheorem{theorem}{Theorem}

\newproclaim{condition}{Condition}

\newcommand{\bY}{{\mathbf Y}}
\newcommand{\balpha}{{\bolds{\alpha}}}
\newcommand{\bpi}{{{\pi}}}

\makeatother

\begin{document}
\begin{frontmatter}

\title{Profile control charts based on nonparametric $\bolds{L}$-1
regression methods\thanksref{T1}}
\runtitle{Profile charts with nonparametric $L$-1 regression}

\begin{aug}
\author[A]{\fnms{Ying} \snm{Wei}\corref{}\thanksref{t2}\ead[label=e1]{yw2148@columbia.edu}},
\author[B]{\fnms{Zhibiao} \snm{Zhao}\thanksref{t3}\ead[label=e2]{zuz13@stat.psu.edu}}
\and
\author[B]{\fnms{Dennis K. J.} \snm{Lin}\ead[label=e3]{DKL5@psu.edu}}
\runauthor{Y. Wei, Z. Zhao and D. K. J. Lin}
\affiliation{Columbia University, Pennsylvania State University
and~Pennsylvania~State~University}
\address[A]{Department of Biostatistics \\
Columbia University \\
722 West 168th St. \\
New York, New York 10032\\
USA\\
\printead{e1}}
\address[B]{Department of Statistics \\
Pennsylvania State University\\
University Park, Pennsylvania 16802\\
USA\\
\printead{e2}\\
\hphantom{E-mail: }\printead*{e3}} 
\end{aug}

\thankstext{T1}{The content is
solely the responsibility of the authors and does not necessarily
represent the official views of the NIDA or the NIH.}

\thankstext{t2}{Supported by NSF Grant DMS-09-06568
and a career award from the NIEHS Center for
Environmental Health in Northern Manhattan (ES-009089).}

\thankstext{t3}{Supported in part by NIDA Grant P50-DA10075-15.}

\received{\smonth{1} \syear{2011}}
\revised{\smonth{7} \syear{2011}}

%
\begin{abstract}
Classical statistical process control often relies on univariate
characteristics. In many contemporary applications, however, the
quality of products must be characterized by some functional relation
between a response variable and its explanatory variables. Monitoring
such functional profiles has been a rapidly growing field due to
increasing demands. This paper develops a novel nonparametric $L$-1
location-scale model to screen the shapes of profiles. The model is
built on three basic elements: location shifts, local shape
distortions, and overall shape deviations, which are quantified by
three individual metrics. The proposed approach is applied to the
previously analyzed vertical density profile data, leading to some
interesting insights.
\end{abstract}

%
\begin{keyword}
\kwd{Functional data}
\kwd{$L$-1 regression}
\kwd{nonparametric methods}
\kwd{profile control charts}.
\end{keyword}

\end{frontmatter}

\section{Introduction}\label{secintro}

Since its initial introduction by Shewart in the 1920s, statistical
process control (SPC) has received increasing attention from both
academia and industry. Traditional SPC often utilizes a single metric
(e.g., mass or length) to characterize products under inspection. For
such metrics, lower and upper control
limits are then estimated from manufacturing data. For example, such
limits could be defined by the
mean plus and minus three standard deviations. If its metric falls out
of the control limits,
it could imply a potential change in the underlying
distribution (for stability),
or a product could be rejected due
to a potential quality deficiency (for conventional quality control).

In traditional SPC, the quality of a product is often assumed to be
adequately characterized by univariate characteristics or certain
metrics. In recent years, however, there have been increasing needs for
profile control
charts for which the responses are no longer single measurements but,
rather, functions of one or several covariates $X$. Traditional methods
handling univariate or multivariate charts are no longer applicable to
such functional profile responses [\citet{Woo07}]. Several new
approaches have been subsequently proposed. One basic class of methods
assumes a linear association between a response $Y$ and its covariate
$X$ and then constructs the corresponding control charts on the intercept,
slope and variance. When linearity is not an option, nonlinear models
are usually introduced. The control charts are then built upon one or
several model coefficients. \citet{JenBirWoo08},
\citet{JenBir09}
further consider parametric (linear and nonlinear) mixed effect models
to take into account intraprofile correlations. \citet{Tsu10} further consider nonparametric mixed effects modeling Phase~II
profile monitoring.

In many applications, it is difficult to know a priori the shape of a
response profile. Inappropriate shape assumptions can lead to
substantial estimation bias. To overcome this difficulty,
\citet{ReiSar06}, \citet{JeoLuWan06}, \citet{DinZenZho06},
\citet{ZhoSunShi07} and \citet{ChiPigSim09} explore nonparametric wavelet
models and construct control charts based on a portion of the wavelet
coefficients. Since the screening method therein relies only on major
wavelet coefficients, any deviations of the other coefficients may be
undetectable. \citet{ZouTsuWan08} also explore an alternative
nonparametric approach for profile monitoring in which the measures
within each profile are assumed to be independent.

The above approaches always project the profile information onto a set
of parameters, while a more natural method is to utilize the entire
information of the target profiles. We hence propose
estimating \textit{a reference profile} based on Phase~I
data and then monitoring the deviations of individual profiles in
Phase~II from the reference one. The basic
statistical tool we use to model Phase~I data is \mbox{$L$-1}
regression assuming a nonparametric location-scale model with
a general class of error structure.
Compared with traditional $L$-2 regression, $L$-1
regression is more robust against outliers and heavy-tailed
distributions. We
refer to \citet{Koe05} for an
extensive exposition on the properties of $L$-1 regression. Based on
the estimated model in Phase~I,
we propose three deviation measures to monitor individual Phase~II
profiles in
overall location shift, local failure and overall shape deviation. The control
limits of the three deviation metrics are determined based on
the empirical estimation of their asymptotic distributions.

\begin{figure}

\includegraphics{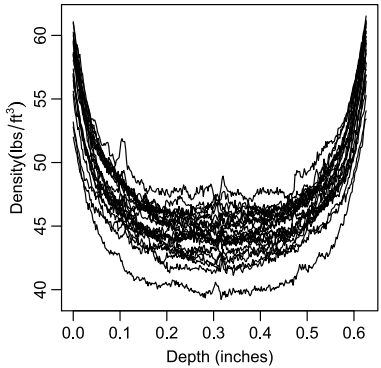}

\caption{Walker--Wright data.}\label{figfig1}
\end{figure}

As an illustrative example, we apply the proposed method to the
vertical density
profile (VDP) data of \citet{WalWri02}. In that study, the
manufacturers of engineered wood boards
are very concerned about fiberboard density, which
determines the fiberboard's strength and physical properties.
The density ($Y$) is read by a
profilometer, a laser device measuring densities at equispaced points
($X$) along a designated vertical line.
Each resulting profile consists of 314 density measurements, and the
distance between two consecutive measures is 0.002 inch. Figure \ref
{figfig1} provides a visual illustration of the data set.
These curves are clearly nonlinear, and thus new technologies
to monitor these curves are desirable.

The rest of our paper is organized as follows. Section \ref{sec2}
introduces a nonparametric location-scale $L$-1 regression with
a generalized error structure, as well as the estimation
procedure. We then
describe in detail how to construct profile control charts
utilizing nonparametric $L$-1 regression. The screening is
based on three deviation measures, whose asymptotic
properties are established. The proposed method is then applied
to the VDP data and illustrated in Section \ref{sec31}. Section
\ref{sec32} compares the proposed method to alternative approaches, and Section
\ref{sec33} provides a
numerical investigation. Section \ref{sec4} presents concluding remarks and a
discussion. The theoretical regularity conditions for
the proposed method are summarized in the \hyperref[app]{Appendix}, while the
proof of the main theorem is presented in the supplementary materials.

\section{Proposed methodology}\label{sec2}

Two phases, often denoted by Phases~I and~II, are typically involved in
SPC [\citet{Woo07}]. The goal in Phase~I is to
construct the control limits which determine if a process has been in
control over the period of time. Phase~II then applies these limits to
detect a potential change in the underlying
distribution (for stability). This paper follows the same convention.

\subsection{Modeling Phase~\textup{I} profiles}

\subsubsection{Representation of Phase~\textup{I} profiles based on a family of
nonparametric location-scale models}

Suppose there exist $n$ independent Phase~I profiles, $\{\bY_i=
(Y_{i,1}, Y_{i,2},\ldots, Y_{i,m_i}), i = 1,\ldots, n\}$, where $m_i$
denotes the number of elements. Let $x_{i,j}$ represent the location
where $Y_{i,j}$ is taken. For instance, if $\bY_i$ is a time sequence,
then $x_{i,j}$ can be the underlying measurement time. Given this
notation, we assume that Phase~I profiles follow a nonparametric
location-scale model
%
\begin{equation}\label{eqmodel}
Y_{i,j}=\delta_i + \mu(x_{i,j})+s(x_{i,j})e_{i,j}, \qquad  1\le j\le
m_i, 1\le i\le n.
\end{equation}
Here we define $\delta_i = \mbox{median}(\bY_i)$ as the marginal
median of the $i$th profile, and view it as the profile center. We also
assume that, for each profile $i$, the error process
$\{e_{i,j}\}_{j\in\N}$ in (\ref{eqmodel}) is an independent
copy from a stationary process with a sufficiently general
dependence structure (as specified in Condition \ref{condc}
in the \hyperref[app]{Appendix}). Such a proposed error structure is well suited for
control chart profiles, mainly because of two factors: (1) it allows
for high correlations induced by dense measurements, differentiating it
from typical longitudinal data settings, and (2) unlike classical time
sequences, it allows the dependence of both left and right neighboring
measurements.
The details of the error structure are discussed in the \hyperref[app]{Appendix}. We
further assume that median$(e_{i,j})=0$ and
median$(|e_{i,j}|)=1$.
Under these assumptions,~$\mu(x)$ is the conditional median of a
centered profile $[\bY(x)- \mbox{median}\{\bY(x)\}]$ given the
location $x$, where $\bY(x)$ is a random profile satisfying model
(\ref{eqmodel}). It represents the standard shape of a normative
centered response profile, and we call it the \textit{reference profile}.
The function $s(x)$ is the conditional median absolute deviation (MAD)
of the centered profile given the location $x$. It measures the extent
to which a normative profile can deviate from the reference
profile at a given location $x$, and we call this the \textit
{reference deviation} function. Hence, we decompose the profiles into
three domains: center, shape, and variability.

\subsubsection{Stepwise estimation}
The key components of model (\ref{eqmodel}) are the profile centers
$\delta_i$, a reference profile $\mu(x)$ and a reference deviation
function~$s(x)$. These can be estimated sequentially as follows.

\paragraph*{Step 1: Estimation of $\delta_i$}

In a simpler case, where all the profiles are measured on a set of
fixed evenly-spaced locations with $m_i \equiv m$, the profile-specific
centers, $\delta_i$, can be estimated by taking the sample median over
the observed $\{Y_{i,j}, j =
1,\ldots, m\}$ for each profile. This type of profiling typically occurs
in manufacturing studies such as the VDP profiles introduced earlier.

In more general applications, the number of measurements
each sequence contains, $m_i$, can vary across profiles, the locations $x_{i,j}$
can be unevenly spaced, and their spacing can also
vary across profiles. To handle such varying location profiles, we can
assume that
each measurement time/location is a random draw from an
underlying distribution $F(x)$, and the observed locations $x_{i,j}$ are
the order statistics of $m_i$ random draws. Letting $X$ be the random
variable following
the distribution $F(x)$, we then
define the profile center as $\delta_i = \arg\min_\theta
E_{Y_i(X)}|Y_i(X) - \theta|$, where $Y_i(x)$ is the underlying\vadjust{\goodbreak}
profile such that $Y_{i,j} = Y_i(x_{i,j})$. Note that since
$E_{Y_i(X)}|Y_i(X) - \theta| = E_X[E_{Y_i(x)}\{|Y_i(x) -
\theta| \vert X=x\}]$, the individual center $\delta_i$ can be
estimated by minimizing the sample objective function $\sum_j
|Y_{i,j} -\theta|\hat{f}(x_{i,j})$, where
$\hat{f}(x_{i,j})$ is the density of $X$ at location $x_{i,j}$ and
can be
estimated from the pooled sample $x_{i,j}$ (over both $i$ and $j$). When
$x_{i,j}$ are evenly spaced,
$f(x)$ is a uniform density, that is, $f(x_{i,j})=f(x_{i,j'})$ for any
$j\not= j'$. Hence, the center can be simply estimated as the sample median.

\paragraph*{\texorpdfstring{Step 2: Estimation of $\mu(x)$}{Step 2: Estimation of $mu(x)$}}
A kernel-based estimation procedure is employed and the algorithm
details are as follows.
Under the specified error structure and other mild conditions (as
listed in the \hyperref[app]{Appendix}),
the resulting estimated functions are uniformly consistent and
asymptotically normal.

Let $K_{b_n}(u)=K(u/b_n)$ be a nonnegative kernel function with
bandwidth $b_n>0$ that satisfies $\int_{\R} K(u) \,d u=1$. We propose
the following least absolute deviation (LAD) estimation for the median
function $\mu(x)$:
%
\begin{equation}\label{eqmuest}
\hat\mu_{b_n}(x) = \argmin_{\theta}\sum_{i=1}^{n} \sum
_{j=1}^{m_i} |Y_{i,j}-\delta_i-\theta| K_{b_n}(x_{i,j}-x).
\end{equation}
The estimation equation (\ref{eqmuest}) is a locally constant type. One
can also extend it to a locally linear estimate, as in
\citet{FanGij96}, without much technical difficulty. We settle on
the locally constant approach mainly for computational simplicity.

Following Theorem 1 of \citet{ZhaWeiLin}, $\{\hat\mu_{b_n}(x)-\mu
(x)\}$ contains a bias term of order $O(b^2_n)$. To remove the bias, we
adopt a corrective jackknife estimator [\citet{WuZha07}]:
%
\begin{equation}\label{eqmbias}
\tilde\mu_{b_n}(x)=2\hat\mu_{b_n}(x)-\hat\mu_{\sqrt{2}b_n}(x).
\end{equation}
The bias-corrected estimator $\tilde{\mu}_{b_n}$ is uniformly
consistent and normally distributed for any $x$ asymptotically
[\citet{ZhaWeiLin}].

\paragraph*{Step 3: Estimation of $s(x)$}
The reference deviation function $s(x)$ is then estimated from the
residuals in the preceding step.
Notice that $\mbox{median}(|e_{i,j}|)=1$ entails $\mbox
{median}(|Y_{i,j}-\delta_i-\mu(x)| \vert x_{i,j}=x) =s(x)$.
Therefore, we propose the following median quantile estimate of $s(x)$:
%
\begin{equation}\label{eqsest}
\hat{s}_{h_n}(x) = \argmin_{\theta}\sum_{i=1}^{n} \sum_{j=1}^{m_i}
\bigl| |Y_{i,j} -\delta_i-\tilde\mu_{b_n}(x)|- \theta\bigr|
K_{h_n}(x_{i,j}-x),
\end{equation}
where $h_n>0$ is another bandwidth and $\tilde\mu_{b_n}(x)$ is a
bias-corrected jackknife
estimator. Similarly, we construct a bias-corrected jackknife estimator
of~$s(x)$ with
%
\begin{equation}\label{eqvbias}
\tilde{s}_{h_n}(x)=2\hat{s}_{h_n}(x)-\hat{s}_{\sqrt{2}h_n}(x),
\end{equation}
which is uniformly consistent and asymptotically normal, as shown in
\citet{ZhaWeiLin}. This stepwise estimation is also used in
\citet{He97} under the constraint that both $\mu(x)$ and $s(x)$
are linear.

\paragraph*{Bandwidth selection}

To implement the proposed methods, one needs to select the proper
bandwidths $b_n$ and $h_n$. Popular choices include plug-in methods,
cross-validation (CV) and generalized CV methods. Originally developed
for independent data, CV methods often tend to undersmooth correlated
data [\citet{OpsWanYan01}]. However, in the absence of universally
efficient alternatives, the CV methods are still among the most widely
used [\citet{FanYao03}, \citet{LiRac07}]. This paper also
adopts the CV approach.

The basic idea is to leave one profile out and fit the model using
the remaining profiles. We then choose the optimal bandwidth that
minimizes the prediction error. Classical
CV methods deal with quadratic losses, whereas here we work with
the $L$-1 loss penalty. Hence, we propose selecting the bandwidth $b_n$
by minimizing
the following modified CV criterion, that is,
%
\begin{equation}\label{eqcrossva}
b^*_n = \argmin_{b_n} \sum^n_{i=1} \sum^{m_i}_{j=1} |Y_{i,j}-\delta
_i-\tilde\mu_{b_n,-i}(x_{i,j})|,
\end{equation}
where $\tilde\mu_{b_n,-i}$ is the bias-corrected jackknife
estimator of $\mu$ based on all but the $i$th profile.
Similarly, we choose $h_n$ with
%
\begin{equation}\label{eqcrossvb}
h^*_n = \argmin_{h_n} \sum^n_{i=1} \sum^{m_i}_{j=1}
\bigl||Y_{i,j}-\delta_i-\tilde\mu_{b_n}|-\tilde{s}_{h_n,-i}(x_{i,j})
\bigr|,
\end{equation}
where $\tilde{s}_{h_n,-i}$ is defined in the same way as $\tilde\mu
_{b_n,-i}$.

\paragraph*{Alternative smoothing approaches}
Other nonparametric estimation methods, such as smoothing splines,
wavelets and normalized B-splines, can also be used to
estimate $\mu(x)$ and $s(x)$, respectively, in steps 2 and 3.
In general, the estimated $\mu(x)$ and $s(x)$ using other smoothing techniques
are also consistent, although their limiting distributions need to be
investigated separately. The proposed CV criterion can also
be adapted to choose other smoothing parameters. We refer to
\citet{OpsWanYan01} for a~detailed comparison of various
smoothing methods.

\subsection{Construct profile control charts}\label{secnonpest}

Suppose $\{(x_l,Y_l), l=1,2,\ldots,m^*\}$ is a new profile from Phase~II, where $m^*$ is the number
of measurements of the new profile. We would like to test
whether it is different from the Phase~I profiles.
Specifically, we are interested in testing the null
hypothesis,
\[
H_0\dvtx   \mbox{The new profile comes from the same profile population,}
\]
against the alternative hypothesis $H_a$, that the new profile
comes from a~different population.

\subsubsection*{Three deviation measures and their control limits} A new
profile can differ from the reference profiles in two ways: through a
vertical shift or a~shape change.\vadjust{\goodbreak} We propose three deviation measures:
one for the first type and two for the latter. The profile control
charts consist of the screening thresholds of the three deviation measures.

To monitor the vertical outliers, we first estimate the center of the
new profile, and denote it as
$\delta^*$. We then standardize it by
\[
D = |\delta^* - \hat{\mu}_\delta|/\hat{s}_\delta,
\]
where $\hat{\mu}_\delta$ and $\hat{s}_{\delta}$ are,
respectively, the sample median and MAD of the Phase~I profile centers
$\delta_i$'s. The deviation score $D$ then provides a relative ranking
of the new profile, from the inside to the outside, with respect to the
Phase~I profiles. To determine the screening threshold for $D$, we can
generate its reference distribution by the empirical (or bootstrap)
distribution of the $d_i= |\delta_i - \hat{\mu}_\delta
|/\hat{s}_\delta$. Consequently, we can use the $(1-\alpha)$th
upper quantile of the $d_i$'s as the control limit and denote it
$c^{(0)}(\alpha)$. Here $\alpha$ is the significance level and will
be determined at a later step.
If the deviation score~$D$ exceeds $c^{(0)}(\alpha)$, then the profile
will be singled out due to its outlying location relative to the Phase~I profiles.

It is more challenging to screen shape deviations. We first center the
profile by $\{Y_l^* = Y_l -\delta^*, l = 1,\ldots, m^*\}$.
The center of the new profile $Y_l^*$ is then zero. By construction,
the reference profile $\mu(x)$ is also centered at zero, which means
there are no systemic distances between the two profiles after the
centering step. Consequently, the main differences between $Y_l^*$ and~$\mu(x)$ are only due to their different shapes.

Recall that $\tilde\mu_{b_n}$ and $\tilde{s}_{h_n}$ are the bias-corrected
estimates of $\mu(x)$ and $s(x)$. We define
%
\begin{equation}\label{eqhatej}
\hat{e}_l:= \frac{Y_l^*-\tilde\mu_{b_n}(x_l)}{\tilde{s}_{h_n}(x_l)},
\qquad 1\le l\le m^*,
\end{equation}
which measures the relative deviation of $Y_l^*$ from the estimated
$\tilde\mu_{b_n}(x_l)$
given the estimated scale function $\tilde{s}_{h_n}(x_l)$. We thus
consider the
following deviation measures:
%
\begin{equation}\label{eqt2}
T^{(1)} = \max_{1\le l\le m^*} |\hat{e}_l|  \quad\mbox{and}\quad
T^{(2)} = \sum^{m^*}_{l=1} |\hat{e}_l|.
\end{equation}
The first statistic $T^{(1)}$ measures the maximal local shape
deviation of the new profile from the estimated reference profile,
while the second score
$T^{(2)}$ measures its cumulative shape deviation. The two scores
complement each other, since one monitors the overall shape change
while the other monitors local perturbations. Together, the scores
provide a
comprehensive monitoring of shapes. \citet{Gar97}
also consider
estimating a ``reference'' surface, using the sum of the residuals to
detect unusual signals.

To determine the screening thresholds for the two measures, we
first need to derive the distributions of $T^{(1)}$ and
$T^{(2)}$ under the null hypothesis. Letting $e_l = \{Y_l -\delta-
\mu(x_l)\}/s(x_l)$ be the error of the new profile, the theorem
cited below establishes the asymptotic distributions of
$T^{(1)}$ and $T^{(2)}$. Here~$T^{(1)}$ has an asymptotic
extreme value distribution [\citet{Gal87}] due to its maximum
structure, while $T^{(2)}$ is asymptotically normally
distributed.
%
%
%
%
\begin{theorem}\label{thmextreme}
Let $N_n$ be as in Condition \ref{conreg} (in the \hyperref[app]{Appendix}).
Define
\[
\Xi_n = b^4_n + h^4_n + \frac{(\log N_n)^{3/2}}{(N_nb_n)^{1/2}}
+ \frac{(\log N_n)^{3/2}}{(N_nh_n)^{1/2}}.
\]
Under Conditions \ref{condc}--\ref{conregularity} (in the \hyperref[app]{Appendix}), the following
statements hold:

\begin{longlist}
\item
Under $H_0$, as $n \to\infty, m^*\to\infty$,
\[
\bigl[T^{(1)}-\beta_{m^*}\bigr]/\gamma_{m^*} \Rightarrow Z,
\]
where $Z$
has the extreme value distribution $F$, and
$ (\gamma_{m^*},\beta_{m^*})_{{m^*}\in\N}$ is a nonrandom
sequence with $\gamma_{m^*} \downarrow0$, such that
$\p\{{\max_{1\le l \le m^*}} |e_l| \le\gamma_{m^*} x +
\beta_{m^*} \} = F(x)$ for all continuity points $x$ of $F$ and
$\Xi_n \beta_{m^*}/\gamma_{m^*}\to0$.

\item Let $e_0$ be an independent and identically distributed (i.i.d.)
copy of~$e_l$ and Condition \ref{condc} (in the \hyperref[app]{Appendix}) holds with
$q=2$. Assume that
the density function of $e_0$ is bounded, and further assume that $\Xi
_n m^{*3/4} \to0$ as
$n,\allowbreak m^*\to\infty$. Then, under $H_0$, as \mbox{$n,m^*\to\infty$},
\[
\frac{T^{(2)} - m^* \mu}{\sqrt{m^*}} \Rightarrow N(0,\sigma^2),
\]
where
%
\begin{equation}\label{eqthem1a}
\mu=\E(|e_0|)  \quad\mbox{and}\quad
\sigma^2 = \operatorname{Var} (|e_0|)
+ 2\sum^\infty_{l=1} \operatorname{Cov}(|e_0|,|e_l|)<\infty.
\end{equation}
\end{longlist}
\end{theorem}

The conditions for Theorem \ref{thmextreme} are summarized in the
\hyperref[app]{Appendix}, and these conditions are rather common in practice. The
proofs are presented in the supplementary materials [\citet{ZhaWeiLin}].

The limiting behavior of $T^{(1)}$ is substantially determined by
the distribution and dependence characteristics of the underlying
$e_l$. Therefore, it is quite challenging to obtain
the critical value (i.e., screening threshold) of~$T^{(1)}$ by using
Theorem~\ref{thmextreme} directly. For $T^{(2)}$, on the other hand,
although Theorem~\ref{thmextreme} guarantees its normal limiting
distribution through (ii), estimating $\mu$ and $\sigma^2$ can be
computationally complicated. Hence, we propose estimating the control
limits numerically based on the Phase~I data. Specifically, we
calculate the shape deviation
scores as in (\ref{eqt2}) for individual Phase~I profiles that
have measurements on $\{x_l, l= 1,\ldots, m^*\}$ and denote them
$T_{i,1}$ and $T_{i,2}$, respectively. The screening thresholds
are then the $(1-\alpha)$th quantiles of $T_{i,1}$ and $T_{i,2}$.
We\vadjust{\goodbreak}
denote the two
screening thresholds by $c^{(1)}(\alpha)$ and
$c^{(2)}(\alpha)$, respectively. The control limits can be obtained
by bootstrapping the Phase~I profiles and using the $(1-\alpha)$th bootstrap
quantiles as the desired screening thresholds.

The derived control limits are valid due to the following reasons: (1)
Despite the difficulties of obtaining asymptotic critical values
directly, Theorem~\ref{thmextreme} establishes the fact that, under
$H_0$, both statistics $T^{(1)}$ and $T^{(2)}$ converge to certain
stable limiting distributions as the number of Phase~I profiles goes to
infinity. (2)~Assuming that the functions $\mu(\cdot)$ and $s(\cdot)$
are known, then, under $H_0$, the distribution of $e_l$ of the new
profile is the same as that of~$e_{i,l}$ from the Phase~I
profiles,\vspace*{1pt} where $e_{i,l} = \{Y_{i,l} -
\mu(x_{i,l})\}/s(x_{i,l})$. Hence, the limiting distributions of
$T^{(1)}$ and $T^{(2)}$ can be well approximated by the empirical
distribution of ${\max_l}|e_{i,l}|$'s and $\sum_l|e_{i,l}|$'s,
respectively, with sufficiently large $n$. Due to the uniform
convergence of $\tilde{\mu}_{b_n}$ and~$\tilde{s}_{h_n}$, the
distribution of~$\hat{e}_{i,l}$ is uniformly and sufficiently close to
that of~$e_{i,l}$, with large enough $n$ and sufficiently small $b_n$
and $h_n$. Combining the above facts, one can generate the reference
distribution of $T^{(1)}$ and $T^{(2)}$ by the empirical (or bootstrap)
distributions of $T_{i,1}$ and $T_{i,2}$.

%

\subsubsection*{Determining the significance level $\alpha$}
In the proceeding steps, we leave the significance level $\alpha$
unspecified and write the three screening thresholds~$c^{(0)}(\alpha
)$, $c^{(1)}(\alpha)$ and $c^{(2)}(\alpha)$ as functions of $\alpha
$. We now choose $\alpha$ such that
%
\begin{eqnarray}\label{eqsig}
\\[-6pt]
\alpha^* = \max_\alpha \Biggl\{\alpha\dvtx \sum_{i=1}^n\max\bigl\{
\mathbf{1}_{\{d_i >c^{(0)}(\alpha)\} },
\mathbf{1}_{\{T_{i,1} >c^{(1)}(\alpha)\}}, \mathbf{1}_{\{
T_{i,2}>c^{(2)}(\alpha)\}}\bigr\}< n \alpha_0
\Biggr\},\nonumber\hspace*{-20pt}
\end{eqnarray}
where $\alpha_0$ is the desired overall significance level. Following
the definition above, a profile will be considered as an outlier if it
appears unusual in any of the three domains. And $\alpha^*$ is the
largest value that ensures that the probability of falsely detecting a
normative profile is less than $\alpha_0$.

\subsubsection*{Summary of the screening procedure} Suppose $(x_l, Y_l)$ is
a new profile. Then the screening consists of the following three steps:
\begin{longlist}
\item Center the profile by its median
$\delta^*$ and calculate its relative vertical deviation by
\[
D =|\delta^*-\hat{\mu}_\delta|/\hat{s}_\delta.
\]
\item Calculate the cumulative and maximal shape
deviation of the centered profile, $Y_l - \delta^*$, with
respect to $\tilde{\mu}_{b_n}(x)$ and
$\tilde{s}_{h_n}(x)$:
\[
T^{(1)}= \max_{1\le l\le m^*} \biggl|
\frac{Y_l-\delta^*-\tilde\mu_{b_n}(x_l)}{\tilde
{s}_{h_n}(x_l)}\biggr|,\qquad
T^{(2)}= \sum^{m^*}_{l=1} \biggl| \frac{Y_l-\delta^*-\tilde
\mu_{b_n}(x_l)}{\tilde{s}_{h_n}(x_l)}\biggr|.
\]
\item If either of $D$, $T^{(1)}$ and $T^{(2)}$
exceeds its corresponding screening threshold,
$c^{(0)}(\alpha^*)$, $c^{(1)}(\alpha^*)$ and
$c^{(2)}(\alpha^*)$, respectively, then the profile
$(x_l, Y_l)$ will be singled out as a potential
outlier.
\end{longlist}

\section{Application}\label{sec3}
\subsection{Profile control chart applied to VDP data}\label{sec31}
\begin{figure}[b]
\begin{tabular}{@{}c@{\quad}c@{}}

\includegraphics{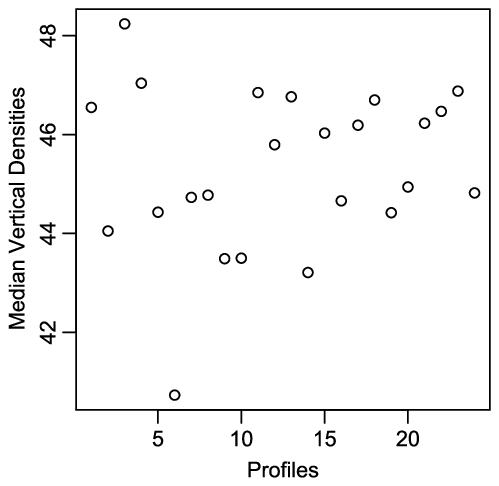}
 & \includegraphics{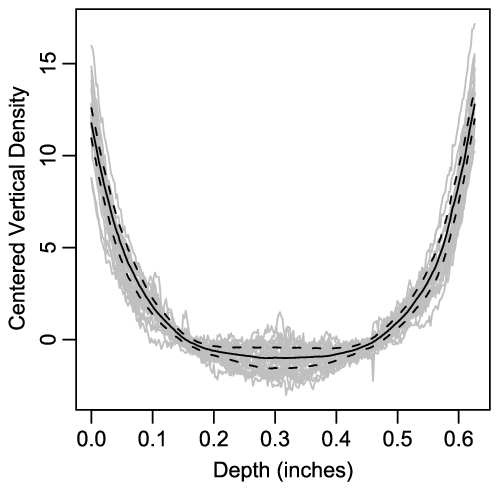}\\
(a) & (b)
\end{tabular}
\caption{VDP profiles. Panel \textup{(a)} presents the centers of the profiles,
while panel \textup{(b)} plots the centered profiles together with their
estimated reference profile (solid curve) and estimated $\mu(x)\pm
s(x)$ (dotted lines).}\label{figthree}
\end{figure}

We now return to the VDP data to illustrate the proposed profile
screening procedure. Recall that the density of the
wood board ($Y$) is measured on a dense sequence of depths~($X$) across
the board.
The VDP data consist of 24 such profiles, and we view them as Phase~I
profiles. We first calculate the center of $Y_i = (Y_{i,1},\ldots,
Y_{i,m})$, which is its overall median $\delta_i = \mbox
{median}(Y_i)$. The profile-specific centers $\delta_i$
represent the overall vertical locations of those profiles and are
presented in
Figure \ref{figthree}(a).

After centering, we estimate the reference profile and reference
deviation functions $\mu(x)$ and $s(x)$ following the proposed
estimation procedure. Based on the CV criterion in (\ref{eqcrossva})
and (\ref{eqcrossvb}), the optimal bandwidths for $b_n$ and $h_n$ are
0.015 and 0.01, respectively. The resulting
$\tilde{\mu}_{b_n^*}(t)$ and $\tilde{s}_{h_n^*}(t)$ are presented in
Figure \ref{figthree}(b) and are depicted as solid (reference
profile) and dashed (deviation profiles) curves.

\begin{figure}

\includegraphics{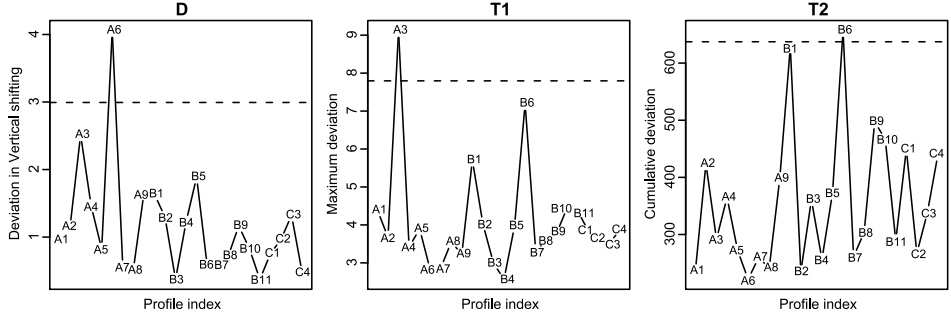}

\caption{Profile charts of relative vertical deviation and
post-centering maximal and cumulative deviations. The dotted lines
are determined control limits.}\label{figcharts}\vspace*{-3pt}
\end{figure}

\begin{figure}[b]
\vspace*{-3pt}
\includegraphics{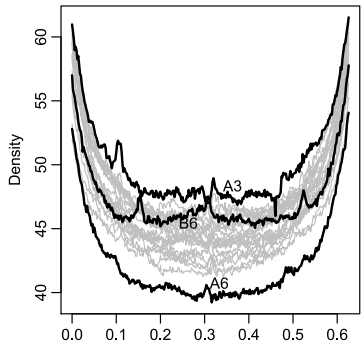}

\caption{Three profiles associated with the largest $D$, $T^{(1)}$ and
$T^{(2)}$.}
\label{figoutliers}
\end{figure}

Following the proposed method to construct the control charts,
we first calculate the relative vertical deviation scores
($d_i$) of the 24 profiles (plotted in the left panel of
Figure \ref{figcharts}). We then calculate the defined
cumulative and maximal deviation scores ($T_{i,1}$ and
$T_{i,2}$) for 24 profiles (plotted in the middle and the
right panels of Figure \ref{figcharts}). The control limits
are determined by assuming a significance level $\alpha=0.03$ for
each deviation score, which yields an overall significance level
$ \alpha_0=0.12\approx3/24$. The resulting screening
thresholds are 2.99, 7.94 and 663.6, respectively. These
control limits are presented by the gray
dotted horizontal lines in Figure \ref{figcharts}. As shown
in Figure~\ref{figcharts}, Profile A6 is associated with the largest
vertical deviation, Profile
A3 is the farthest outlier in the maximal shape deviation, while
Profile B6 has the largest cumulative
shape deviation of all the centered profiles.
Although they are assumed to be
normative profiles, these profiles provide insights on what
the maximum tolerated vertical and shape deviations are. We plot these
curves in their original forms in Figure
\ref{figoutliers}. It is clear that Profile A6 has a lower
vertical density than all the other profiles, Profile B6 is
``too flat'' in the center section compared to the rest
%
of the profiles, while Profile~A3 has stronger local turbulence than
others. Although the VDP data have been used in various studies,
our conclusion provides some new insights into the data set.

\subsection{Comparisons with other approaches}\label{sec32}

The VDP data have been investigated
by many others. Most approaches are based on linear profile
assumptions, as\vadjust{\goodbreak} in \citet{Wooetal04}, \citet{KimMahWoo03},
\citet{Mahetal07}, \citet{KanAlb00} and
\citet{ZhuLin10}. Although those linear models may be
applicable in some situations, they are clearly inappropriate
for the VDP data due to their nonlinear nature. Other nonlinear
profile approaches, on the other hand, tend to be more
complicated, and some are hard to implement in practice. Two
of the most relevant approaches to our work are the bathtub model
proposed by \citet{WilWooBir07} and the $\chi^2$ control
charts of \citet{ZhaAlb09} and \citet{Shietal09}. We elaborate
these authors' approaches below and discuss how they differ from our methods.

\citet{WilWooBir07} fit a bathtub function
\[
f(x_{ij}, \beta)=
\cases{
a_1 (x_{ij}-c)^{b_1}+d, &\quad $x_j > c$; \cr
a_2 (x_{ij}-c)^{b_2}+d, &\quad $x_j \le c$,}
\]
to each profile (for Profiles \#$1, 2,\ldots, 24$), yielding 24
sets of estimates for a six-dimensional vector of parameter
$\beta= ( a_1, a_2, b_1, b_2, c, d)$. The authors then construct (i)
six univariate control charts for each of the parameters
($a_1,\ldots, d$) and (ii) a multivariate $T^2$ control chart for
the vector of $\beta$. Based on those control charts, they
conclude (page 934) that boards \#4, 9, 15, 18 and
24 have outlying profiles. Note that our identified
farthest outlying Profiles A3, A6 and B6 correspond to Profiles
\#3, \#10 and \#15 in their setup. Although only \#15 is
detected by \citet{WilWooBir07}, the authors do point out
(page 935) that Profiles \#6 and \#3 should be
outliers as well, which is consistent with our conclusion.
Since \citet{WilWooBir07} restrict the shapes of the
profiles to a family of bathtub models, the bathtub model may exhibit
a certain lack of fit in some profiles,
which could, in turn, lead to failure in detecting Profiles
\#6 and \#3. Moreover, the bathtub model suffers from an
identifiability issue that affects the control charts based on it. That
is, two distinct sets of parameters can
yield nearly identical bathtub curves.

\citet{ZhaAlb09} assume that all the profiles are measured on a
fixed grid of locations/times. This way, the profiles can be viewed as
long vectors. Consequently, one can construct a $\chi^2$ control chart
based on the individual quadratic distances with respect to an
estimated mean vector ($\mu_s$) and an estimated variance--covariance
matrix ($\Sigma_s$), that is, $\Delta_i=(y_i-\hat{\mu
}_s)^\prime\hat{\Sigma}_s^{-1} (y_i-\hat{\mu}_s)$. When the
profiles are densely measured, the variance--covariance matrix $\Sigma$
can be close to singular, which makes the estimation challenging. Using
this approach, \citet{ZhaAlb09} identify Profi\-les~\#3, 6, 9, 10
and 15 as outliers. This fully covers our findings of Profiles~\#3, 10
and 15. \citet{Shietal09} further apply functional principal
component analysis to individually smoothed profiles and then monitor
potential outliers based on the quadratic distance of the major
principal component scores. Compared to these approaches in
\citet{Shietal09} and \citet{ZhaAlb09}, the proposed charts have the
following two advantages: First, the proposed method has the
flexibility to accommodate random locations, in which case the profiles
can be observed for unevenly spaced and individual sets of locations.
Second, the proposed approach decomposes the potential deviations into
three domains: location, shape, and local disturbance. Consequently,
the screening results provide more information with which to detect
outliers. In this sense, our proposed approaches can be viewed as
``targeted'' screening, compared to these ``generic'' screening
approaches.

In addition, the proposed model has different setup and noise structure
from the nonparametric mixed effect model in
\citet{Tsu10}. Specifically, they assumed the model
$y_{ij}=g(x_{ij})+f_i(x_{ij})+\epsilon_{ij}$,
where $g$ is the population profile function, $f_i$ is the
random-effects term due to the $i$th individual profile, and
$\epsilon_{ij}$'s are i.i.d. random errors with mean 0 and variance~$\sigma^2$.
Essentially, in the proposed model (1), we further write out
the random function $f_i(x_{ij})$ as $\delta_i +
s(x_{i,j})e(x_{i,j})$, where $e(\cdot)$ is a random process with a general
correlation structure. This
specification does not reduce the model flexibility, and makes it
easier to handle the heteroscedasticity and location shifts.


\subsection{Numerical investigation using synthetic VDP profile-like
data}\label{sec33} 
\begin{figure}
\begin{tabular}{cc}

\includegraphics{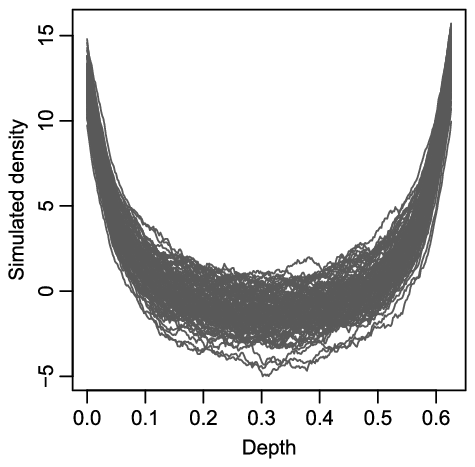}
 & \includegraphics{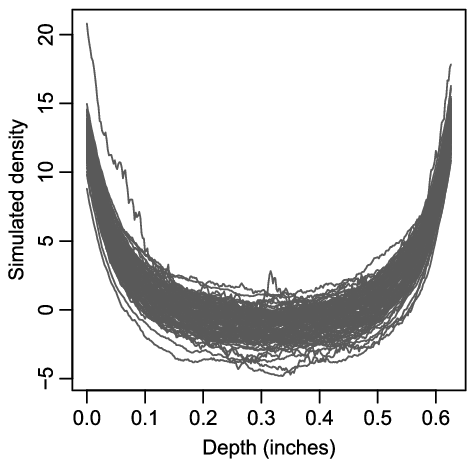}\\
(a) & (b)
\end{tabular}
\caption{Simulated profiles. The profiles in panel \textup{(a)} are generated
from the Gaussian process, while those in panel \textup{(b)} follow a scaled
$t_3$ process.}\label{figsimuldata}
\end{figure}

To investigate the numerical performance of the proposed method,
we generate synthetic
data sets that mimic the VDP data, based on which we evaluate the
screening power of the proposed control charts. To simulate VDP-like
data, we choose the same set of locations ($x$) as in the VDP data,
which range from 0 to 0.626, with a grid length of 0.002. We then
generate 100 individual density profiles at the chosen locations based
on the
following model:
%
\begin{equation} \label{eqsimulmodel}
Y_i(x) = \delta_i + \bpi(x)^\top\balpha_0 + e_i(x),
\end{equation}
where $\bpi(x)$ is an eight-dimensional quadratic B-spline basis
function with internal knots
(0.06, 0.16, 0.31, 0.47, 0.56), which are the 0.1th, 0.25th, 0.5th,
0.75th and 0.9th sample quantiles of the locations $x$ in the VDP data,
respectively. We further assume that $\delta_i$ are i.i.d. random
coefficients that follow a normal distribution $N(0, \sigma_{\delta
}^2)$. Finally, we consider two stochastic processes for the error term
$e_i(x)$: (1) $e_i(x) \sim N(0, \sigma^2)$ and (2)
$e_i(x)$ follows a scaled $t$ distribution with three degrees of
freedom and variance $\sigma^2$. In addition, we assume an
exponentially decay correlation structure for both error processes,
that is,
%
\begin{equation}\label{eqsimulerror}
\mbox{corr}(e_i(x), e_i(x')) = \exp\{-8|x-x'|\}.
\end{equation}
Note that, in both cases, the median $e_i(x) =0$ for all $x$,
hence, $\delta_i$ is the median location by construction, that is,
the center of the profiles. The generated profile $Y_i(x)$ then
follows model (\ref{eqmodel}) with median location (the
center) $\delta_i$, reference function $\mu(x) =
\bpi(x)^\top\balpha_0$ and reference deviation function
$s(x)\equiv1$. For sensible choices of $\balpha_0$,
$\sigma_\delta^2$ and $\sigma^2$, we estimate them from the original VDP
data. Specifically, we regress individual profiles over $\bpi(x)$
using $L$-1 regression and denote the resulting coefficient
estimate $\tilde{\balpha}_i$. We choose $\balpha_0$ by the
sample mean of $\tilde{\balpha}_i$, choose
$\sigma_{\delta}^2$ by
the sample variance of the $\delta_i$ obtained in the preceding section,
and then choose $\sigma$ as the standard deviation of the pooled residuals.
Figure \ref{figsimuldata}(a) displays the generated
%
profiles~$Y_i(x)$ under the two error distributions. We see that the
profiles generated from the $t$ process
have more turbulence than those generated from the Gaussian process.

\subsubsection*{Screening} 
\begin{figure}
\begin{tabular}{@{}c@{}}

\includegraphics{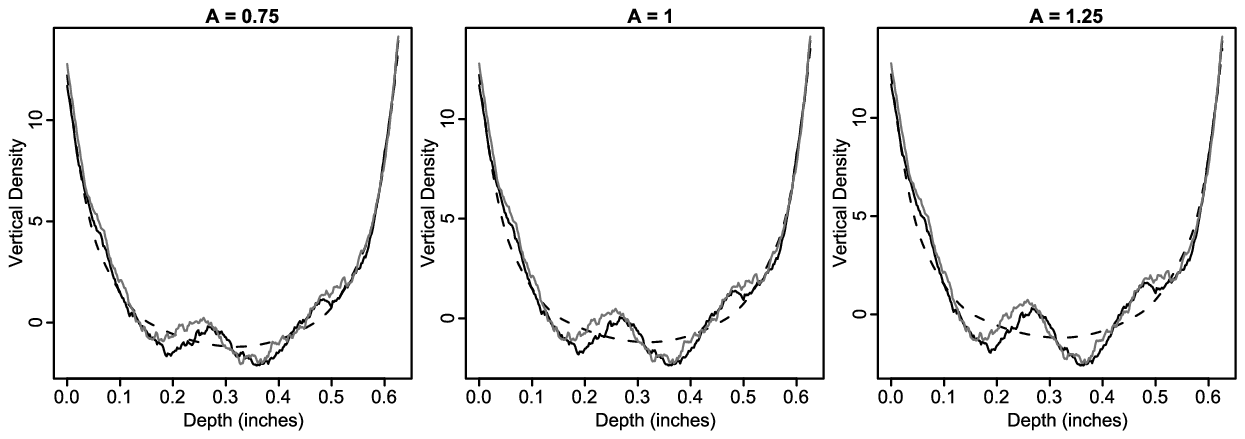}
 \\
(a) \\[6pt]

\includegraphics{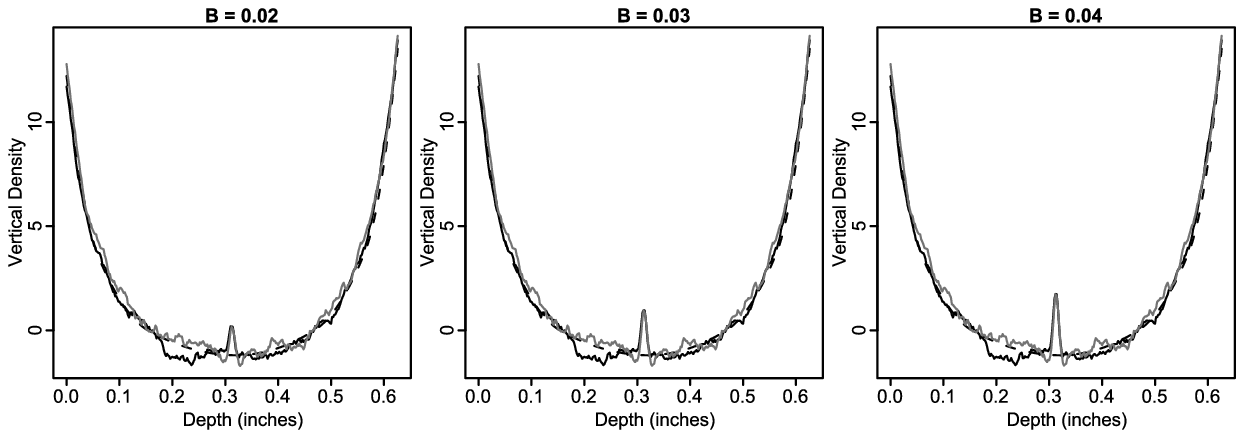}
\\
(b)
\end{tabular}
\caption{Simulated density profiles from the misspecified
models \textup{(a)} and \textup{(b)}. The dotted curve is the true reference profile, the
solid black curves are simulated paths from the $t_3$ process, and the
solid dark gray curves are those generated from the Gaussian
process.}\label{figwrongmodel}
\end{figure}

We next construct the control charts using the
proposed me\-thod. When estimating the reference and deviation functions,
we choose the optimal CV bandwidths $b_n$ and $h_n$ as 0.004 and 0.007,
respectively, for the Gaussian $e(x)$, and 0.01 and 0.007,
respectively, for the $t_3$ distributed~$e(t)$. The three control
limits are obtained assuming the overall significance level of 0.05
following equation (\ref{eqsig}). To investigate the screening power
of the proposed control charts for the two sets of profiles above, we
generate another 100 profiles from the true model (\ref
{eqsimulmodel}) and from each of the following two ``wrong'' models:
\begin{eqnarray*}
\mbox{Model (a)\quad} Y_i(x) &=& a_i + \pi(x)^\top\balpha+
A\sin(10\pi x) + e_i(x), \\
\mbox{Model (b)\quad} Y_i(x) &=& a_i + \pi(x)^\top\balpha+
B\phi(x-0.3)/0.005 + e_i(x),
\end{eqnarray*}
where $a_i$ and $e_i(x)$ follow the same error process in the true
model, either the Gaussian or the $t_3$ distribution, and $\phi(\cdot
)$ is the density function of a~standard normal. Model (a) distorts the
shape of
the profile by adding a~sine curve with the coefficient $A$ determining
its amplitude, while model (b)
introduces local ``spiky'' noise to the
profile. The noise level is determined by the coefficient $B$.
We consider the coefficients $A$ = 0.75, 1 and 1.25 for model~(a),
and $B$ = 0.02, 0.03 and 0.04 for model (b), representing
smaller to larger contamination scales. Figure
\ref{figwrongmodel} displays the simulated profiles from
%
the two misspecified models with the Gaussian and $t_3$ processes,
respectively. As we can see in Figure \ref{figwrongmodel}, when the
coefficients $A$ and $B$ increase, the severity of the noise
also increases. We then apply the proposed screening procedure,
with the resulting proportions of successful identification
presented in Table~\ref{table1}. When the profiles are generated
from the true model, the false discovery rates are 0.05 ($\pm0.02$) and
0.08 ($\pm0.02$), respectively. They are close to their nominal
level 0.05. For misspecified models, the
screening power increases with the amplitude of the noise. In
both cases, we have decent power to detect moderate to larger
profile deviations. When the error $e(t)$ follows the~$t_3$
distribution, the profiles fluctuate more, and, consequently,
the screening power is lower than that of the Gaussian process.

\section{Discussion}\label{sec4}
%

This paper proposes to screen the shape of the profiles
using a nonparametric location-scale $L$-1 model. The basic
idea is to estimate a reference profile curve, based on which
we can rank the differences of individual profiles from the
reference profile. The nonparametric components of the proposed
%
\begin{table}
\tabcolsep=0pt
\caption{The percent of the profiles that exceed the control limits.
Under the true model, type \textup{I} error is
reported. Under models \textup{(a)} and \textup{(b)}, the detection rate is reported}
\label{table1}
\begin{tabular*}{\tablewidth}{@{\extracolsep{\fill}}lccccccc@{}}
\hline
& & \multicolumn{3}{c}{\textbf{Model (a)}} & \multicolumn
{3}{c@{}}{\textbf{Model (b)}}\\[-4pt]
& & \multicolumn{3}{c}{\hrulefill} &
\multicolumn{3}{c@{}}{\hrulefill}\\
& \textbf{True model} &
$\bolds{A = 0.75}$&$\bolds{A = 1.00}$&$\bolds{A = 1.25}$
& $\bolds{B = 0.02}$& $\bolds{B = 0.03}$&$\bolds{B =
0.04}$ \\
\hline
Gaussian $e(t)$& 5\% &45\% &80\% &\hphantom{0}98\% &36\% &82\% &100\% \\
$t_3$ distributed $e(t)$& 8\% & 21\% &62\% &100\% & 12\% &44\% &\hphantom{0}95\% \\
\hline
\end{tabular*}
\end{table}
model provide sufficient flexibility to capture the shape of the
profiles. In addition, as inherited from $L$-1 regression, we do not
assume any specific distributions for the profiles, and the resulting
estimates are fairly robust against the heavy-tail distributions and
data contamination.

\subsection*{$L$-1 versus $L$-2 screenings}

The proposed screening procedure relies on \mbox{$L$-1}
regression. However, it remains valid if one replaces $L$-1
regression with $L$-2 regression (least-squares
regression), which is more commonly used for control charts.
Specifically, to achieve $L$-2 screening, we redefine $\xi_i$
in model~(\ref{eqmodel}) as the conditional mean of the $i$th
profile, and define~$\mu(x)$ and~$s(x)$, respectively, as the
conditional mean and standard deviation functions of $Y_i(x) -
\delta_i$. The centers $\delta_i$ can be estimated by the sample mean of
the $i$th profile, and the functions $\mu(x)$ and $s(x)$ can be
estimated using kernel smoothing to replace the absolute value
functions in (\ref{eqmuest}) and (\ref{eqsest}) with square
functions. Consequently, we define the deviation score $d_i$ by
the standardized~$\delta_i$, using its mean and standard deviation,
and keep the deviation scores~$T_1$ and $T_2$ in the same form,
except that $\mu(x)$ and $s(x)$ are now the conditional mean and
standard deviation functions. The $L$-2 screening could be
more efficient and effective for normative profiles, but the
$L$-1 screening is known to be more robust if the Phase~I
profiles contain potential outliers. Practitioners can choose according
to their specific needs.

Profile monitoring is a relatively new area, but it is growing rapidly, as
indicated by increasing numbers of practical applications.
The proposed methods are shown to be favorable when applied to
the previously analyzed VDP data. They can be further
generalized to reach a wider range of applications. First of all,
we rank the shape deviation of an individual profile from the reference
profile based on its largest residual and cumulative residuals. The asymptotic
distributions of the resulting deviation scores are studied.
Depending on the application, alternative deviation scores can be used.
For example, instead of the largest residual, we can use the 90th
percentile of the
residuals. The properties of other deviation scores would be of
research interest.
Second, the proposed charts consist of three metrics, monitoring
deviations in location, shape and local disturbances, respectively. If
one type of deviation is of major concern, one could focus on one
metric (more targeted screening) to achieve better screening power.
Third, in some applications, the data may exhibit specific features.
Incorporating these special features into the estimation can further
improve efficiency. For example, suppose that the VDP profile is
assumed to be symmetric around the midpoint $c_x$. If we assume that
the location function $\mu(x)$ is a smooth differentiable function, a
more efficient way to estimate it is to regress $y_{i,j}$ over the
transformed location $x_{i,j}^* = |x_{i,j}-c_x |$, that is, estimate
the conditional median function $\mbox{median}(y_{i,j}|x_{i,j}^*) =
g(x_{i,j}^*)$ with the constraint $g'(0)=0$. Consequently, the location
function $\mu(x)$ equals $g(|x-c_x|)$. Fourth, we assume in the
proposed model that the errors are in a stationary random sequence. The
developed theories can be further generalized to incorporate
more general error sequences. For example, $e_{i,j}$ could be a linear
combination of
$k$ stationary error sequences. Fifth, the control limits are estimated
from data, and hence subject to certain estimation errors
[\citet{Jenetal06}]. When there are insufficient Phase~I profiles, some
corrections may be needed to ensure good properties of the proposed
control charts.
Finally, note that
we use Phase~I profiles to generate the reference distributions
of the shape deviation scores. This approach implicitly assumes that
there are a sufficient number of Phase~I profiles that have been
measured at
similar locations as the profile to be screened. In the case of sparse profile
data with irregular spacing, we may not have sufficient Phase~I
profiles to ensure a stable estimation of the screening thresholds.
Further research is needed to deal with such sparse profile data.

\begin{appendix}\label{app}

\section*{\texorpdfstring{Appendix: Conditions of Theorem \lowercase{\protect\ref{thmextreme}}}{Appendix: Conditions of Theorem 1}}

%

Recall that, for a given $i$, $\{e_{i,j}\}_{1\le j\le
m_i}$ is the error process of the $i$th
profile. We assume that there exist an independent process
$\{e_{\ell}\}_{\ell\in\Z}$, such that the $i$th error profile
$\{e_{i,j}\}_{1\le j\le
m_i}$ can be viewed as one realization of the process $\{e_{\ell}\}
_{1\le\ell\le m_i}$.
We assume that
$\{e_\ell\}_{\ell\in\Z}$ has the representation
%
\begin{equation}\label{eqerror}
e_\ell=G(\varepsilon_\ell,\varepsilon_{\ell\pm1}, \varepsilon
_{\ell\pm2},\ldots),
\end{equation}
where $\varepsilon_\ell, \ell\in\Z$, are i.i.d. random vectors,
and $G$ is a measurable function such that $e_\ell$ is well
defined. The representation (\ref{eqerror}) can be viewed as
an input-output system with
$(\varepsilon_\ell,\varepsilon_{\ell\pm1},
\varepsilon_{\ell\pm2},\ldots)$, $G$ and $e_\ell$ being the
input, transform or filter, and output, respectively. This
representation allows for noncausal models, and is hence
particularly useful for our applications, which do not have a
time structure.

For $q>0$ and a random variable $e$ we denote by
$\|e\|_q=[\E(|e|^q)]^{1/q}$ the $L_q$ norm. We further assume the following:
%
\begin{condition}[(Dependence condition)]\label{condc}
Let $e_0$ be as in (\ref{eqerror}), and
$(\varepsilon'_\ell)_{\ell\in\Z}$ be an i.i.d. copy of
$(\varepsilon_\ell)_{\ell\in\Z}$. There exist $q>0$ and
$\rho\in(0,1)$ such that for all $k\in\N$,
%
\begin{eqnarray}\label{eqgmc}
\|e_0-e_0(k)\|_q=O(\rho^k)\hspace*{80pt}\nonumber\\[-8pt]\\[-8pt]
&&\eqntext{\mbox{where }
e_0(k) = G\bigl(\varepsilon_0,\varepsilon_{\pm1},\ldots,\varepsilon
_{\pm k}, \varepsilon'_{\pm(k+1)}, \varepsilon'_{\pm(k+2)},\ldots\bigr).}
\end{eqnarray}
\end{condition}

In (\ref{eqgmc}), $e_0(k)$ can be viewed as a coupling process
of $e_0$ with $\varepsilon_r$ being coupled by the i.i.d. copy
$\varepsilon'_r$ for $|r|\ge(k+1)$, while keeping the nearest
$2k+1$ innovations $\varepsilon_r$ with $|r|\le k$. In
particular, if $e_0$ does not depend on
$(\varepsilon_r)_{|r|\ge(k+1)}$, then $e_0(k)=e_0$. Thus,
$\|e_0-e_0(k)\|_q$ can be viewed as the impact of
$(\varepsilon_r)_{|r|\ge(k+1)}$ on~$e_0$. Intuitively, it is
reasonable to expect that measurements sufficiently far away
would have negligible impact. In particular, condition
(\ref{eqgmc}) states that the impact decays exponentially
as the location space $k$ increases. As shown in Lemma
1 of \citet{ZhaWeiLin}, (\ref{eqgmc}) implies that the
correlation between $e_0$ and $e_\ell$ decays exponentially
as $\ell$ increases.
%
\begin{condition}[(Location condition)]\label{conreg}
The set of measurement locations $\{x_{i,j},
1\le j\le m_i, 1\le i\le n\}$ is asymptotically uniformly dense
in $[a,b]$. Specifically, let $a=\tilde{x}_0 < \tilde{x}_1
<\cdots<\tilde{x}_{N_n} <\tilde{x}_{N_n+1}= b$ be the ordered
locations, where $N_n=m_1+\cdots+m_n$ is the total number of
measurements. We assume that
${\max_{0\le k\le N_n}}|\tilde{x}_{k+1}-\tilde{x}_k - \frac
{b-a}{N_n}|=O(N^{-2}_n)$.
\end{condition}
%
\begin{condition}[(Kernel condition)]\label{conlc}
Let ${\mathcal K}_\omega, \omega>0$, be the set of kernels which are
bounded, symmetric, and have bounded support\vspace*{1pt}
$[-\omega,\omega]$ with~boun\-ded derivative. Let $\varphi_K=\int _\R
K^2(u)\,d u$ and $\psi_K=\int_\R u^2 K(u) \,d u / 2, \mbox{$K\in{\mathcal
K}_\omega$}$. The kernel $K\in{\mathcal K}_\omega$.
\end{condition}
%
\begin{condition}[(Smoothness condition)]\label{conregularity}
Denote by $F_e$ and $f_e=F'_e$, respectively, the
distribution and density functions of $e_0$ in
(\ref{eqerror}). Assume~$\mu,\allowbreak s\in{\mathcal C}^4([a,b]),
\inf_{s\in[a,b]} s(x)>0, f_e\in{\mathcal C}^4(\R), f_e(0) > 0,
f_e(1)+f_e(-1)>0$. He\-re~${\mathcal C}^4(S)$ is the set of 4th order
continuously differentiable functions on set~$S$.
\end{condition}

Condition \ref{conreg} says that the pooled locations of
measurements should be reasonably uniformly dense, even though
each single profile may only contain sparse measurements. The
remaining two conditions are also fairly general, and commonly
assumed in kernel smoothing. Under Conditions \ref{condc}--\ref{conregularity} and $\Xi_n\to
0$ (see Theorem \ref{thmextreme} for definition), the location
and scale function estimates, $\tilde\mu_{b_n}$ and
$\tilde{s}_{h_n}$, are uniformly consistent.
\end{appendix}



\begin{supplement}[id=suppA]
\stitle{Proof of Theorem \ref{thmextreme}}
\slink[doi]{10.1214/11-AOAS501SUPP} 
\slink[url]{http://lib.stat.cmu.edu/aoas/501/supplement.pdf}
\sdatatype{.pdf}
\sdescription{The technical proof of Theorem \ref{thmextreme} is provided in the
supplementary material.}
\end{supplement}


\printaddresses

\end{document}